\documentclass[twocolumn,superscriptaddress, notitlepage,amssymb, nobibnotes,nofootinbib,longbibliography]{revtex4-1}

\usepackage{float}
\usepackage{amsmath}
\usepackage{xcolor}
\usepackage{enumitem}
\usepackage{graphicx}
\usepackage{dcolumn}
\usepackage{bm}
\usepackage{upgreek}
\usepackage{amssymb}
\usepackage{ulem}
\usepackage{svg} 

\usepackage{xr}
\usepackage{wrapfig,booktabs}
\usepackage{tikz}
\usetikzlibrary{arrows}
\usepackage{pgfplots}
\usepackage{setspace}
\usepackage{textcomp}
\usepackage{tabularx}
\usepackage{color}
\definecolor{Blue}{rgb}{0,0.08,0.9}
\definecolor{Black}{rgb}{0,0,0}
\definecolor{Red}{rgb}{0.9,0,0.08}
\usepackage[colorlinks=true, allcolors=blue]{hyperref}

\pgfplotsset{compat=1.18}
\begin{document}

\title{Quantum optomechanical control of long-lived bulk acoustic phonons}


\author{Hilel Hagai Diamandi} 
\email{hagai.diamandi@yale.edu}
\affiliation{Department of Applied Physics and Yale Quantum Institute, Yale University, New Haven, CT, USA, 06511} 
\author{Yizhi Luo} 
\affiliation{Department of Applied Physics and Yale Quantum Institute, Yale University, New Haven, CT, USA, 06511} 
\author{David Mason} 
\affiliation{Department of Applied Physics and Yale Quantum Institute, Yale University, New Haven, CT, USA, 06511} 
\author{Tevfik Bulent Kanmaz} 
\affiliation{Department of Applied Physics and Yale Quantum Institute, Yale University, New Haven, CT, USA, 06511} 
\author{Sayan Ghosh}
\affiliation{Department of Applied Physics and Yale Quantum Institute, Yale University, New Haven, CT, USA, 06511} 
\author{Margaret Pavlovich} 
\affiliation{Department of Applied Physics and Yale Quantum Institute, Yale University, New Haven, CT, USA, 06511} 
\author{Taekwan Yoon}
\affiliation{Department of Applied Physics and Yale Quantum Institute, Yale University, New Haven, CT, USA, 06511} 
\author{Ryan Behunin} 
\affiliation{Department of Applied Physics and Material Science and Center for Material Interfaces in Research and Applications, Northern Arizona University, Flagstaff, AZ, USA, 86011}
\author{Shruti Puri}
\affiliation{Department of Applied Physics and Yale Quantum Institute, Yale University, New Haven, CT, USA, 06511} 
\author{Jack G.E. Harris}
\affiliation{Department of Applied Physics and Yale Quantum Institute, Yale University, New Haven, CT, USA, 06511} 
\affiliation{Department of Physics, Yale University, New Haven, CT, USA, 06511} 
\author{Peter T. Rakich} 
\email{peter.rakich@yale.edu}
\affiliation{Department of Applied Physics and Yale Quantum Institute, Yale University, New Haven, CT, USA, 06511}

\begin{abstract}
\noindent \textbf{Abstract:} 
High-fidelity quantum optomechanical control of a mechanical oscillator requires the ability to perform efficient, low-noise operations on long-lived phononic excitations.
Microfabricated high-overtone bulk acoustic wave resonators (\textmu HBARs) have been shown to support high-frequency ($>10$ GHz) mechanical modes with exceptionally long coherence times ($>1.5$~ms), making them a compelling resource for quantum optomechanical experiments.
In this paper, we demonstrate a new optomechanical system that permits quantum optomechanical control of individual high-coherence phonon modes supported by such \textmu HBARs for the first time.
We use this system to perform laser cooling of such ultra-massive (7.5~\textmu g) high frequency (12.6 GHz) phonon modes from an occupation of $\sim$22 to fewer than 0.4 phonons, corresponding to laser-based ground-state cooling of the most massive mechanical object to date.
Through these laser cooling experiments, no absorption-induced heating is observed, demonstrating the resilience of the \textmu HBAR against parasitic heating.
The unique features of such \textmu HBARs make them promising as the basis for a new class of quantum optomechanical systems that offer enhanced robustness to decoherence, necessary for efficient, low-noise photon-phonon conversion. 
\end{abstract}

\flushbottom
\maketitle

\section*{Introduction}
\label{Intro}

Through the use of engineerable photon-phonon interactions, cavity optomechanical techniques allow efficient control of phonons using light~\cite{aspelmeyer2014cavity,eichenfield2009optomechanical,tomes2009photonic,bahl2011stimulated,hu2013optomechanical,yang2015laser,li2021cavity,lake2021processing}, transforming them into a versatile quantum resource~\cite{tian2010optical,stannigel2010optomechanical,stannigel2012optomechanical,hill2012coherent,weaver2017coherent,ren_two-dimensional_2020,xia2023entanglement}.
Efficient photon-phonon coupling permits the use of quantum optics methods to control such solid-state excitations, allowing for the manipulation and storage of nonclassical states using light~\cite{chan2011laser,fiore2011storing,peterson2016laser,clarke2018growing,qiu2020laser,mirhosseini2020superconducting,behunin2023harnessing}.
In this context, long-lived phonons are advantageous as they may permit numerous quantum operations within the phonon’s coherence time, enabling a new class of high-performance quantum sensors, transducers, and memories~\cite{wallucks2020quantum,fiaschi2021optomechanical}.
To date, some of the most advanced demonstrations of quantum optomechanical control have been enabled by nano-scale phononic crystal resonators that produce long-lived gigahertz-frequency phonon modes, yielding ground-state operation at millikelvin temperatures \cite{maccabe_nano-acoustic_2020,barzanjeh2022optomechanics,mayor2024two}.

While such nanomechanical systems offer large optomechanical coupling rates and long phonon lifetimes, their small size makes them susceptible to phonon dephasing and parasitic optical heating~\cite{meenehan2015pulsed,maccabe_nano-acoustic_2020,barzanjeh2022optomechanics,mayor2024two}.
Hence, new strategies are required to enable efficient, low-noise photon-phonon conversion for efficient state swaps and rapid ground state initialization in cavity optomechanics.
Quantum optomechanical systems based on bulk acoustic wave resonators have the potential to overcome these challenges.
Microfabricated high-overtone bulk acoustic resonators (\textmu HBARs), which are Fabry-Perot resonators for bulk acoustic phonons, support stable, high Q-factor ($>10^8$) phonon modes. 
Such systems can be made from wide-bandgap crystals that have very low optical absorption \cite{pinnow1973development} and high thermal conductivity \cite{touloukian1970experimental}, making them robust to unwanted optical heating \cite{kharel2022multimode}.
Such \textmu HBARs support phonons with large motional masses ($\sim10$\textmu g) and small surface participation, producing highly coherent phonon modes~\cite{galliou2013extremely,kharel2018ultra} that exhibit excellent noise characteristics using qubit-based studies~\cite{chu2017quantum,chu2018creation,bourhill_generation_2020,von2022parity,schrinski2023macroscopic,bild2023schrodinger,marti2024quantum}.
In recent studies, such \textmu HBARs have been shown to support high frequency ($> 10$~GHz) phonons with long ($> 1$~ms) coherence times{~\cite{royce2024}}.
However, quantum optomechanical control of such \textmu HBARs presents many challenges, as these systems produce a densely packed spectrum of high Q-factor phonon modes with small zero-point coupling rates ($\sim 10$ Hz)~\cite{renninger2018bulk,kharel2019high}.
Therefore, practical quantum optomechanical systems require new strategies to both selectively address individual phonon modes of the \textmu HBAR and to enhance optomechanical coupling rates.

Previous studies have shown that cavity optomechanical techniques can be employed to boost coupling rates to bulk acoustic phonons~\cite{kharel2019high, kharel2022multimode,doeleman2023brillouin}.
These earlier proof-of-principle experiments utilized resonantly enhanced Brillouin scattering to access phonon modes of unstable (flat-flat) HBAR resonators~\cite{kharel2019high, kharel2022multimode}, and demonstrate low-noise ground state operation at milikelvin temperatures \cite{doeleman2023brillouin}.
However, diffraction losses within these unstable resonators constrain the achievable phonon Q-factors ($<10^5$), limiting the utility of these systems for quantum optomechanical applications due to their short phonon lifetimes ($\sim 1$~\textmu s) and the relatively high optical powers ($\sim1$mW) required to reach unity cooperativity. 
Alternatively, an optomechanical system that can make use of the stable, ultra-high Q-factor modes of the \textmu HBARs could significantly increase cooperativities, while transforming these highly-coherent phonons into a valuable quantum resource.

Here, we use a new Brillouin-based cavity optomechanical system to realize ground-state cooling of high-frequency (12.6 GHz) phonon modes supported by a \textmu HBAR, enabling quantum optomechanical control of such ultra-massive, highly coherent mechanical oscillators.
To boost the optomechanical coupling, the \textmu HBAR is integrated within an optical Fabry-Perot (FP) resonator to realize a triply resonant optomechanical system that enables resonantly enhanced inter-modal Brillouin scattering~\cite{behunin2023harnessing,kharel2019high,kharel2022multimode,yoon2023simultaneous,doeleman2023brillouin}.
Selective coupling to an individual phonon mode is achieved using a combination of phase matching, mode matching, and spectral filtering within the cavity optomechanical system.
Such resonantly enhanced coupling rates combined with high phonon Q-factors supported by the \textmu HBAR enable unity cooperativity at optical powers as low as 11~\textmu W.
We use this system to demonstrate laser cooling of these ultra-massive (7.5 \textmu g) \textmu HBAR phonon modes from an average thermal occupation of 22.4 to the ground state ($<$ 0.4), corresponding to a cooperativity of 60, using modest ($<2$ mW) powers.
This is the most massive object that has been laser-cooled to its ground state, providing new avenues for fundamental tests of quantum mechanics in the macroscopic regime~\cite{galliou2013extremely,goryachev2014gravitational,lo2016acoustic,neuhaus2021laser,schrinski2023macroscopic,agafonova2024lasercooling1milligramtorsional}.
{Moreover, these results and techniques pave the way for a new class of robust quantum optomechancial systems that could permit low-noise operations with near-unity efficiency for future quantum experiments.}
\begin{figure*}
     \centering
     \includegraphics[width=\textwidth]{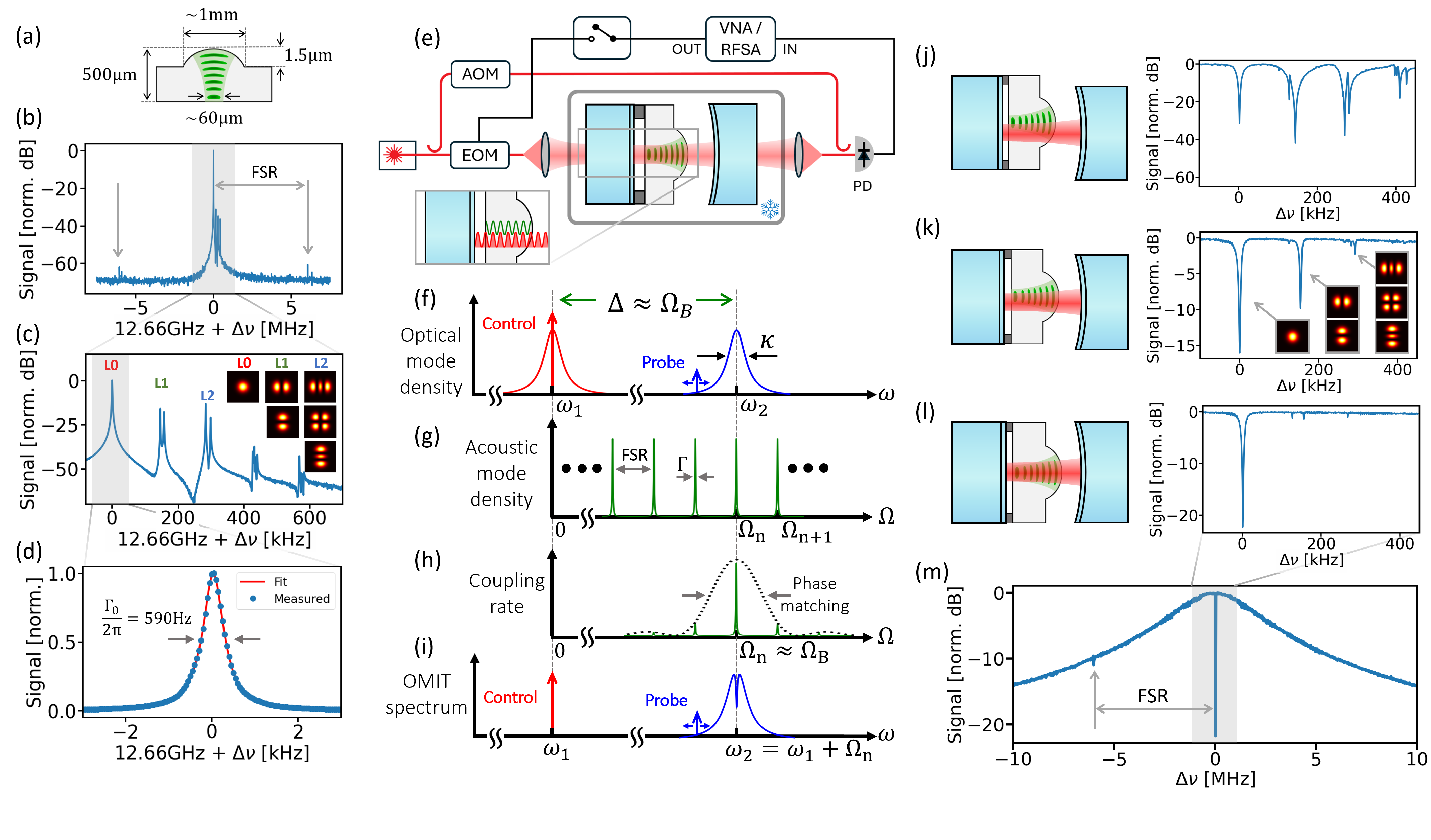}
     \vspace{-10pt}
    \caption{\textbf{Design and characterization of Brillouin-based optomechanical system.} \textbf{(a)} Schematic showing dimensions of micro-fabricated High-overtone Bulk Acoustic Resonator (\textmu HBAR).
    \textbf{(b)} Optically driven acoustic response of the \textmu HBAR, showing repeated acoustic resonances with a 6.04 MHz acoustic FSR.  
    \textbf{(c)} High-resolution measurement of the driven acoustic response, showing Hermite-Gaussian (HG) mode families (L0, L1, L2) with 140~kHz transverse mode spacing.
    \textbf{(d)} Driven acoustic response of the fundamental (L0) acoustic mode showing a linewidth of $\Gamma_0/2\pi = 590$ Hz at 7~K.
    \textbf{(e)} Schematic of \textmu HBAR-based cavity optomechanical system and measurement apparatus; \textmu HBAR is positioned within an optical Fabry-Perot (FP) that is designed to produce acousto-optic mode matching and cooled to 13.6~K.  
    Light in the upper arm of apparatus is frequency-shifted by an Acousto-Optic-Modulator (AOM) to create a local oscillator (LO) for heterodyne detection.
    Light in the lower arm is modulated by an electro-optic modulator (EOM), generating a probe tone used for OMIT and OMIA measurements.
    Light is transmitted through the cavity using fiber collimators, mixed with the LO, and detected with a high-speed photo-detector;
   detected signals are analyzed with a vector network analyzer (VNA) during OMIT/OMIA measurements (switch closed) or an RF spectrum analyzer (RFSA) during spontaneous measurements (switch open).
    \textbf{(Inset)} Phase-matched optomechanical coupling occurs when the spatial periodicity produced by the interference of two adjacent optical modes (red) matches the spatial period of the phonon mode (green), and their frequency detuning matches the phonon frequency.
    Panels \textbf{(f)-(i)} illustrate how phase-matching and spectral filtering are used to realize single mode coupling. Within the optical FP mode spectrum (f), two modes with frequency separation matching the Brillouin frequency ($\Omega_B$) are used to realize resonantly enhanced optomechanical scattering. (g) shows a uniform phonon mode-spacing produced by the fundamental \textmu HBAR mode. (h) illustrates a modulation of the optomechanical single photon coupling rate ($g_o$) produced by phase-matching in the vicinity of the Brillouin frequency. (i) is the single-mode OMIT spectrum resulting from the combination of phase-matching and filtering produced by this system.   
  Panels \textbf{(j)-(l)} show OMIT traces for different alignments of the \textmu HBAR within the optical cavity, from misaligned (j) to optimally aligned (l); optimal alignment (l) yields single-mode (L0) coupling with $>$ 20 dB suppression of higher order (L1, L2, etc) modes.
    \textbf{(m)} Broad OMIT scan for the optimally aligned case of panel (l) showing single-mode coupling with the fundamental acoustic mode (compare to panel (l)), and very weak residual coupling to adjacent fundamental mode at $\Delta \nu=-6$~MHz (see Supplement Sec. VIII). 
    }
     \label{fig:HBAR}
     \vspace{-15pt}
\end{figure*}

\section*{Experimental Results}
\label{Methods}
\subsection*{The Cavity Optomechanical System}
\label{sub:HBAR}

In what follows we build our cavity optomechanical system around the \textmu HBAR seen in Fig.~\ref{fig:HBAR}a.
The \textmu HBAR is a Fabry-Perot (FP) resonator for bulk acoustic phonons, supporting a series of high Q-factor Gaussian modes. 
The \textmu HBAR is created by shaping the surfaces of a 500 \textmu m-thick z-cut $\alpha$ quartz substrate into a plano-convex geometry using a reflow-based fabrication process~\cite{kharel2018ultra}.
The \textmu HBAR has a 100~mm radius of curvature and cavity length of 500~\textmu m, forming a stable Gaussian-beam resonator for longitudinal bulk acoustic phonons.
This design produces fundamental longitudinal acoustic modes with a waist radius of 31~\textmu m at 12~GHz frequencies, corresponding to a motional mass of 7.5~\textmu g~\cite{kharel2018ultra,doeleman2023brillouin}. 
At 1550 nm wavelengths, mode matching with a Gaussian laser beam permits mode-selective coupling to phonon modes near the Brillouin frequency (12.65 GHz), as dictated by the phase-matching condition for acousto-optic Bragg scattering \cite{renninger2018bulk}.

The \textmu HBAR's phonon mode spectrum is independently measured at cryogenic temperatures using noninvasive laser-based Brillouin spectroscopy of the type described in ~\cite{royce2024}.
As seen from the measured phonon mode spectra (Fig.~\ref{fig:HBAR} b-c), this resonator supports families of longitudinal modes, with a free-spectral range (FSR) of 6.04~MHz, and a transverse mode spacing of 140~kHz. 
Measurements of the fundamental phonon mode at T = 7 K (Fig.~\ref{fig:HBAR}d) reveal a phonon Q-factor of 21 million (590 Hz linewidth) at 12.66~GHz, corresponding to an $f-Q$ product of $2.7 \times 10^{17}$ Hz, a key figure of merit that characterizes decoupling of a resonator from its thermal environment~\cite{aspelmeyer2014cavity}.
Similar \textmu HBAR resonators have yielded record $f-Q$ products of $1.8\times10^{18}$ Hz, corresponding to a 1.8 ms coherence time \cite{royce2024}.
Hence, with optomechanical control, such long-lived phonons could become a compelling quantum resource. 
However, conventional cavity optomechanical techniques based on sideband-resolved coupling become impractical due to the high frequencies of the Brillouin-active phonon mode ($\sim 12$~GHz) and relatively small zero-point coupling rates ($g_0/2\pi < 20$ Hz) produced by photo-elastic coupling to these massive phonon modes.

To address this challenge, we use resonant enhancement of the Brillouin interaction to dramatically boost the optomechanical coupling rates~\cite{kharel2019high}. 
The \textmu HBAR is placed inside a high-finesse ($F\cong3000$) plano-concave optical FP resonator (Fig.~\ref{fig:HBAR}e) with a mode spacing ($\Delta \omega = \omega_2-\omega_1$) that matches the frequency of the Brillouin-active phonon mode ($\Omega_n$) falling within the Brillouin phase-matching bandwidth.
This permits the incident pump photons to be scattered between the optical cavity modes by the Brillouin-active phonon mode through resonant inter-modal scattering (Fig.~\ref{fig:HBAR}f-i).
When pumping either optical mode, resonant enhancement of the intra-cavity photon number ($n_c$) dramatically increases the optomechanical coupling rate ($g = g_0 \sqrt{n_c}$) to the macroscopic  \textmu HBAR resonator phonon modes~\cite{kharel2019high}. 
We will show that such resonantly enhanced coupling schemes enable cooperativity of $C>1$ at microwatt power levels, and efficient quantum control of these massive, high Q-factor phonon modes.
The mode spectrum of the optical FP resonator is designed to create a significant sideband asymmetry, as needed for quantum control of the \textmu HBAR phonon modes. 
Weak intra-cavity reflections created by the quartz \textmu HBAR surfaces result in non-uniform optical mode spacing~\cite{kharel2019high}. This non-uniform mode spacing is used to shape the spectrum and control scattering processes~\cite{kharel2019high}.
By selecting two optical modes ($\omega_1,\omega_2$) within this non-uniform spectrum that satisfy the condition for resonant scattering ($\omega_2-\omega_1 \approx \Omega_B$), with adjacent mode pairs' spacings that are significantly detuned, we inhibit unwanted scattering processes~\cite{kharel2019high}.
This symmetry-breaking results in a large ($>$ 1,000-fold) difference between the Stokes and anti-Stokes (AS) scattering rates, allowing us to virtually eliminate the Stokes or AS interaction by resonantly exciting an appropriately chosen optical mode (see Supplement Sec. III).
However, even when these conditions are satisfied, numerous \textmu HBAR phonon modes can mediate scattering between these optical cavity modes.
As seen in Fig.~\ref{fig:HBAR}h, Brillouin scattering permits coupling to multiple longitudinal mode families ($\Omega_{n-1},\Omega_{n},\Omega_{n+1}$) within the Brillouin phase-matching bandwidth (dashed line), centered around the Brillouin frequency ($\Omega_\mathrm{B}$). 
Hence, control of an individual phonon mode requires new strategies to suppress scattering to unwanted phonon modes.
Selective coupling to a single phonon mode is achieved using phase matching in conjunction with mode-matching, and spectral filtering within the cavity optomechanical system.
By designing the optical cavity to have a linewidth ($\kappa/ 2 \pi \cong 4$~MHz) smaller than the acoustic FSR ($\sim$~6 MHz), we utilize the spectral filtering from the optical cavity to effectively restrict the coupling to an individual mode family. 
However, the use of this triply-resonant coupling scheme necessitates matching of the cavity mode spacing ($\omega_2-\omega_1$) to the frequency of the Brillouin-active phonon mode ($ \Omega_n$) with very high precision ($<$~1~MHz). 
The optical cavity mode spacing is adjusted to match the frequency of the fundamental Gaussian mode of interest using piezo-actuation of the cavity length (see Supplement Sec. I~A,III for details).
When the resonance condition ($\omega_2-\omega_1 = \Omega_n$) is met, this system yields an optomechanically-induced transparency (OMIT) spectrum of the type sketched in Fig.~\ref{fig:HBAR}i~\cite{aspelmeyer2014cavity,kharel2019high}. 


The experimental OMIT spectra of Fig.~\ref{fig:HBAR}j-m are obtained by resonantly pumping the red mode ($\omega_1$) with the control laser while a probe tone, generated by intensity modulation of the control laser, is swept through the blue cavity mode ($\omega_2$) to perform a transmission measurement.
Pound-Drever-Hall (PDH) locking is used to precisely align the control laser frequency to the red mode while the transmitted probe light is measured using heterodyne detection using the apparatus of Fig.~\ref{fig:HBAR}e (see Supplement Sec. I~B for further details). 
The OMIT spectrum of Fig.~\ref{fig:HBAR}m shows that spectral filtering by the cavity mode greatly suppresses coupling to other mode families, revealing an individual OMIT dip (or dips) centered at the frequency of the optical resonance.
However, high-resolution measurements of the OMIT spectrum reveal coupling to the fundamental (L0) as well as unwanted higher-order spatial modes (L1, L2, L3) within a given mode family~(Fig. \ref{fig:HBAR}j).

To minimize coupling to higher-order acoustic modes, we precisely match the optical and acoustic field distributions, allowing for mode-selective coupling to the fundamental (L0) Gaussian mode.
Using an optical cavity length of 12~mm and a 15~mm mirror radius of curvature, we realize a fundamental optical mode with an intensity waist radius (39~\textmu m), closely matching that of the \textmu HBAR acoustic field amplitude (31~\textmu m), as necessary to selectively couple to the fundamental (L0) Gaussian phonon mode.
To precisely align the \textmu HBAR within the optical FP resonator for optimal coupling, the crystal is mounted on a flexure stage that permits fine-tuning of the \textmu HBAR position (see Supplement Sec. II~A for details).
As seen from Fig.~\ref{fig:HBAR}k and Fig.~\ref{fig:HBAR}l, precise trimming of the \textmu HBAR position to optimize the transverse alignment of optical and acoustic modes leads to a significant reduction in coupling to higher-order acoustic modes. 
In the case of optimal alignment (Fig.~\ref{fig:HBAR}l-m), efficient coupling to the fundamental Gaussian mode at 12.607~GHz is achieved, with $>$20~dB of suppression of all other acoustic resonances, enabling single-mode quantum control.

Note that the position of the \textmu HBAR affects the spectrum of the optical cavity modes during this alignment process. Hence, following the repositioning of the \textmu HBAR, a pair of optical modes at a slightly different wavelength fulfills the condition for resonant coupling. 
Since the Brillouin frequency is wavelength dependent, the exact frequency of the phonon modes observed in OMIT spectra Fig.~\ref{fig:HBAR}j-m also varies slightly with each alignment. For this reason, the OMIT spectra are presented as functions of detuning from the fundamental acoustic mode of the relevant mode family. 
(See Supplement Sec. II~B).

\subsection*{Ground-State Cooling of \textmu HBAR Phonons}
In what follows, we use this cavity optomechanical system to perform ground-state cooling. 
We begin by introducing the system Hamiltonian and key quantities to describe the cooling process. 
In the case of optimized single-mode coupling, the interaction Hamiltianan for our system becomes $\smash{\mathcal{H}_{\mathrm{int}} = -\hslash g_0(\hat{a}_1\hat{a}_2^\dag\hat{b}_n + \hat{a}_1^\dag\hat{a}_2\hat{b}_n^\dag)}$ in the rotating wave approximation~\cite{kharel2019high}.
Here, $\hslash$ is Plank's constant; $\hat{a}_{1}$ ($\hat{a}_{2}$) is the annihilation operator for the red (blue) mode with angular frequency $\omega_1$ ($\omega_2$); $\smash{\hat{b}_{n}}$ is annihilation operator phonon mode with angular frequency $\Omega_n$; and $g_0$ is the single-photon coupling rate for resonant inter-modal scattering produced by an individual \textmu HBAR phonon mode~\cite{kharel2019high}.

Resonant pumping of this system produces the familiar beam-splitter and squeezing Hamiltonians. 
When pumping the red mode, the substitution $\hat{a}_1\!\rightarrow\! \alpha_1$ yields a linearized beam-splitter Hamiltonian of the form $\smash{\mathcal{H}_{\mathrm{B.S.}} \!= \!-\hslash \Tilde{g}(\hat{a}_2^\dag\hat{b}_n + \hat{a}_2\hat{b}_n^\dag)}$, where $\Tilde{g} = g_0\alpha_1 $ is the effective coupling rate and $\alpha_1$ is the amplitude of the coherent state in mode $\hat{a}_1$.
Similarly, when pumping the blue mode, the substitution $\hat{a}_2\rightarrow \alpha_2$ yields a squeezing Hamiltonian of the form $\smash{\mathcal{H}_{\mathrm{Sq.}} = -\hslash \Tilde{g}(\hat{a}_1\hat{b}_n + \hat{a}_1^\dag\hat{b}_n^\dag)}$, where $\Tilde{g} = g_0\alpha_2$ is the effective coupling rate and $\alpha_2$ is the amplitude of the coherent state in mode $\hat{a}_2$. 
The cooperativity for this optomechanical system is $C = 4|\Tilde{g}|^2/(\Gamma_0\kappa)= 4|\Tilde{g_0}|^2|\alpha|^2/(\Gamma_0\kappa)$, where $|\alpha|^2$ is the intra-cavity photon number ($n_c$) produced through resonant pumping~\cite{kharel2019high}. 
Hence, resonant enhancement of the photon number ($n_c$) enables much larger effective coupling rates ($\tilde{g}$) and cooperativities than are feasible using conventional sideband-resolved optomechanical pumping schemes \cite{aspelmeyer2014cavity,kharel2019high,kharel2022multimode}.

\begin{figure}
     \centering
     \includegraphics[width = \linewidth]{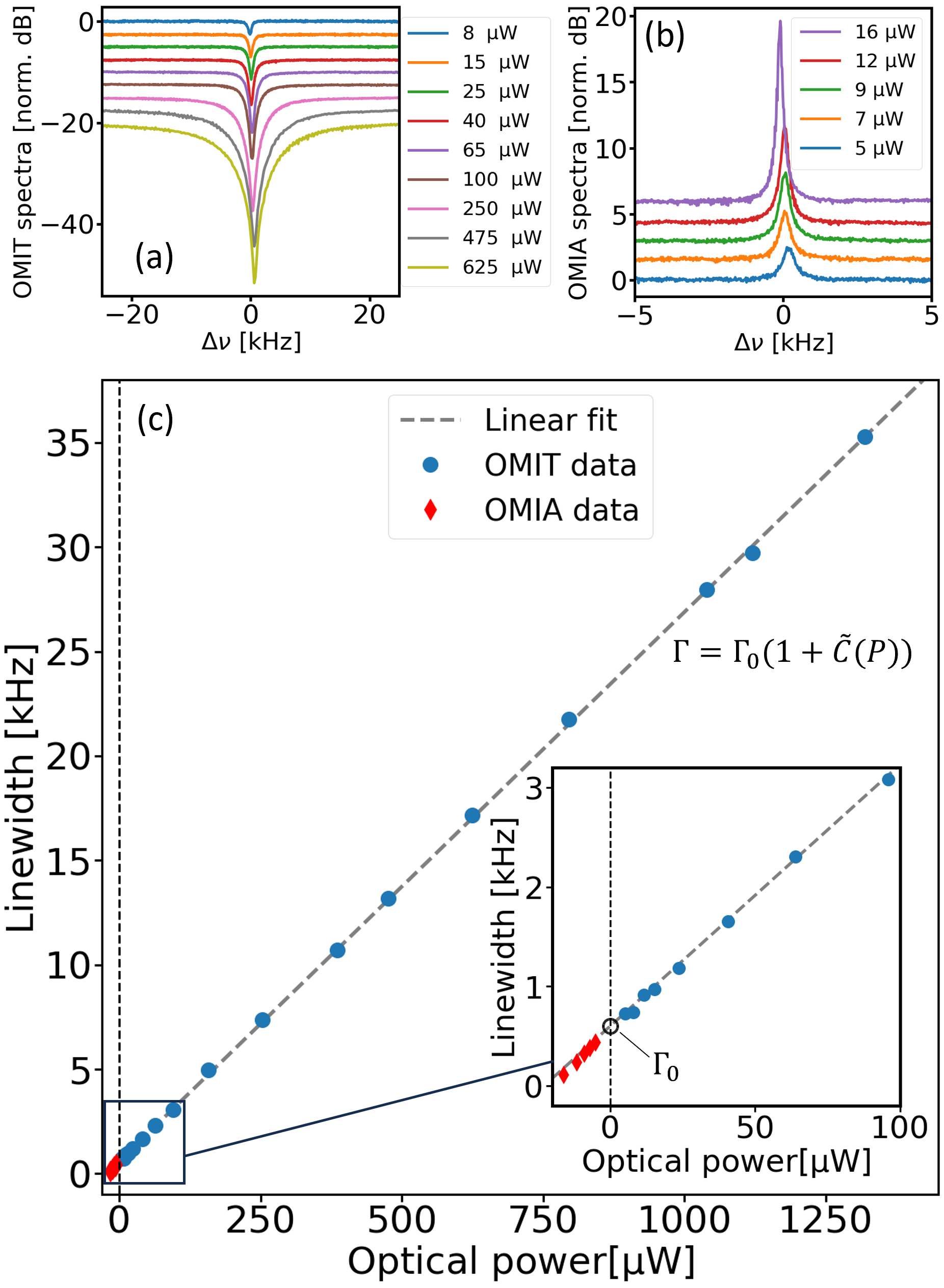}
     \caption{\textbf{OMIT and OMIA measurements.} \textbf{(a)} OMIT traces for several transmitted optical power levels showing broadening of the OMIT dip with increased power (each trace is artificially offset by 2.5 dB for better visibility). \textbf{(b)} OMIA traces for several transmitted optical power levels showing narrowing of the OMIA peak with increased power (each trace is artificially offset by 1.5 dB for better visibility). \textbf{(c)} Each OMIT (OMIA) trace is fit to a Lorentzian and the evaluated linewidths are plotted as a function of transmitted power in blue circles (red diamonds). OMIA data is reflected to negative powers to enable single linear fit (dashed line) of the entire dataset. \textbf{(Inset)} Magnified view of the low power levels region. Black circle marks the fitted fundamental linewidth $\Gamma_0 = 2\pi\times600$ Hz.}
     \label{fig:2}
     \vspace{-15pt}
 \end{figure}

To quantify the coupling rate ($g_0$) and the fundamental linewidth of the phonon mode ($\Gamma_0$) within our cavity optomechanical system, we begin by performing power-dependent OMIT and optomechanically induced amplification (OMIA) measurements. 
The OMIT (OMIA) spectra of Fig.~\ref{fig:2}a-b are obtained by resonantly pumping the red (blue) mode with the control laser while a probe tone, generated by intensity modulation of the control laser, is swept through the blue (red) cavity mode at frequency $\omega_{2}$ ($\omega_{1}$) to perform transmission measurements; during such measurements
{the non-resonant, lower (higher) side-band generated by the modulator has negligible impact, as it is rejected by the optical cavity.}
As before, PDH-locking is used to ensure that the control laser remains on resonance with the appropriate cavity mode during each measurement.
The transmitted optical carrier and probe tone are collected by a collimator and combined with a fixed, frequency-shifted local oscillator (LO) to enable heterodyne measurement of the transmitted probe-wave. 

Next, we analyze these data to extract the parameters of the optomechanical system.
OMIT measurements taken at control laser powers between 7 \textmu W and 1.3 mW are shown in Fig.~\ref{fig:2}a.
The optical power for each trace was measured in transmission to ensure that it is representative of the intracavity photon number. 
(see Supplement Sec. V for further details.)
Panel b of Fig.~\ref{fig:2} shows measured OMIA traces for a series of transmitted control laser power levels below \textit{C}~=~1, which marks the onset of optomechanical self-oscillation.

To extract the effective damping rate of the phonon mode at each optical power, we first fit these experimental traces to the Lorenzian form predicted from the OMIT and OMIA response, yielding the data seen in Fig.~\ref{fig:2}c. 
We then fit these extracted power-dependent linewidths to the theoretical damping rates given by $\Gamma_{\pm}(C) = \Gamma_0 \pm \Gamma_{\mathrm{opt}}=\Gamma_0(1 \pm C)$ predicted for OMIT ($\Gamma_{+}(C)$) and OMIA ($\Gamma_{-}(C)$)~\cite{aspelmeyer2014cavity,kharel2019high}. 
Since $C$ is proportional to the intra-cavity photon number, $\Gamma_{+}(C)$ ($\Gamma_{-}(C)$) increases (decreases) linearly with optical power for the case of OMIT (OMIA) measurements.
To enable a single linear fit for the entire dataset, the OMIA data is reflected to negative power levels, as seen in Fig.~\ref{fig:2}c.
The linear fits of the data reveal a fundamental linewidth of $\Gamma_0 = 2\pi\times600 (\pm30)$~Hz. 
Unity cooperativity (\textit{C}~=~1) is achieved at transmitted power level of $22.8\pm1.2$\textmu W, corresponding to a single-photon coupling rate of $g_0 = 6.08\pm0.03$~Hz.  
Hence, mode-selective coupling to the high Q-factor \textmu HBAR phonon modes permits us to achieve $C=1$ with orders of magnitude smaller optical powers than prior studies \cite{kharel2022multimode}.
(For a complementary dataset demonstrating unity cooperativity at 11 \textmu W transmitted power levels, see Supplement Sec. VIII.)


\begin{figure}
     \centering
     \includegraphics[width = \linewidth]{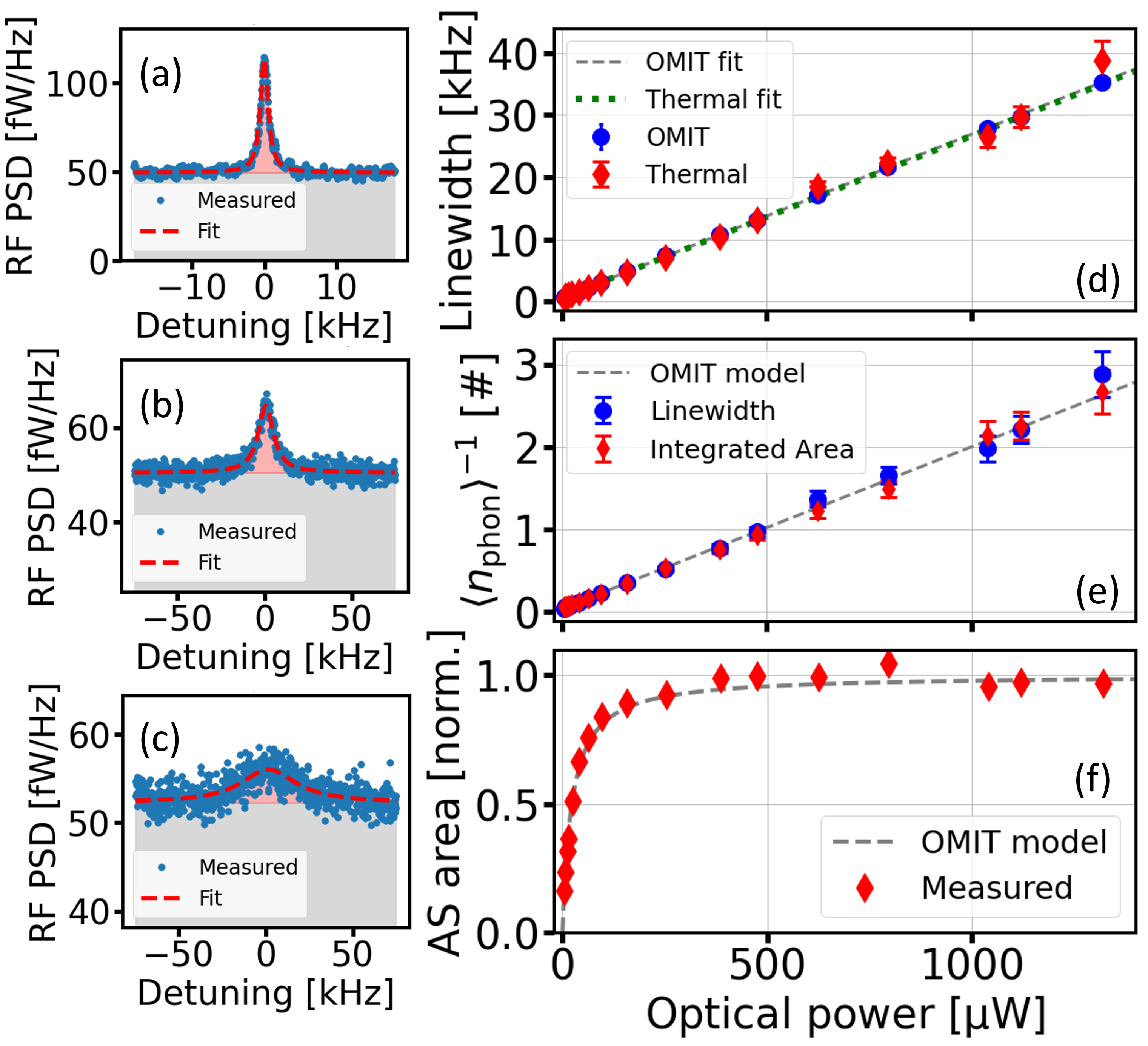}
     \caption{\textbf{Laser cooling measurements.} \textbf{(a)-(c)} Spontaneous measurements of the RF power spectral density (PSD) of the heterodyne-detected Anti-Stokes light scattered by the cooling process for growing control power levels of 24 \textmu W, 386 \textmu W, and 1.3 mW, in panels (a),(b), and (c), respectively. The measured traces are fit to Lorentzians (red dashed lines), and the extracted linewidths and areas (shaded red) are used to evaluate the cooled phonon population. The background levels (shaded gray) match the expected increase due to shot noise, and show no significant heating of the mode by the control laser (see Supplement Sec. IV,VII).
     More spontaneous AS cooling traces can be found in Supplement Sec. VIII.
     \textbf{(d)} Extracted Lorentzian linewidths of spontaneous measurements for different control powers (red diamonds) show excellent agreement with the linear fit and the OMIT measurements (blue circles).  
     \textbf{(e)}
     Starting from $\langle n \rangle = n_{\mathrm{th}}$, the actively cooled steady-state phonon population is deduced from the measured linewidth (blue circles) and area (red diamonds) of each trace. Calculations yield values in good agreement with the analytical model (gray dashed line). The lowest phonon population for transmitted power of 1.3 mW is  $\langle \mathrm{n} \rangle \simeq$  0.36 phonons.  
     \textbf{(f)} Measured (diamonds) normalized areas of the AS traces show good agreement with the OMIT based model of $\tilde{C}(P)/(1+\tilde{C}(P))$.}
     \label{fig:Cooling_exp}
     \vspace{5pt}
 \end{figure}

To perform laser cooling, we resonantly pump the red mode ($\omega_1$) with the control laser while measuring the spontaneous AS light emitted from the blue mode ($\omega_2$); during these measurements, the probe tone is turned off.
The frequency-shifted optical local oscillator is combined with the transmitted light to perform heterodyne spectral measurement of spontaneously scattered AS light.  
These spectra were acquired using a high-speed photodetector and an electrical spectrum analyzer.  
The optical local oscillator is amplified and passed through a high-rejection narrow-bandwidth filter to enable a shot-noise limited detection.
Panels a-c of Fig.~\ref{fig:Cooling_exp} show examples of spontaneously measured AS spectra with increasing control laser powers of 24 \textmu W, 386 \textmu W, and 1.3mW, respectively. The traces are fit to Lorentzian lineshapes (dashed red lines) from which the linewidths and areas (shaded red) are extracted.
For increasing optical powers, the AS linewidth broadens in a manner that is consistent with increasing optical damping of the phonon mode; these measurements also show a change in the peak brightness of the AS Lorenzian spectrum that is consistent with a reduction in phonon population (See Supplement Sec. IV).
The complete dataset of spontaneous AS spectra can be found in Supplement Sec. VIII.

Using these spectra, in conjunction with the measured parameters of the cavity optomechanical system, we quantify the reduction in the phonon population produced by laser cooling.
The linewidths extracted from the AS spectra (diamonds) are plotted in Fig.~\ref{fig:Cooling_exp}d, demonstrating consistency with those extracted from OMIT measurement (circles) at each power level. 
The area of the Lorenzian AS resonance, which is proportional to the emitted AS photon flux, also follows the predicted $\tilde C(P)/(1+\tilde C(P))$ dependence, as seen in Fig.~\ref{fig:Cooling_exp}f. Here, $\tilde C(P) $ is the linear fit of cooperativity as a function of transmitted power, extracted from the OMIT data.

Next, we use these measurements to estimate the phonon occupation number.
Due to the unique features of our system, sideband asymmetry measurements ~\cite{safavi2012observation,purdy2015optomechanical,underwood2015measurement,meenehan2015pulsed,doeleman2023brillouin,mayor2024two} are inaccessible (See Supplement Sec. IV). Therefore, we base the thermometry measurements on the available linewidth and area of the measured anti-Stokes spectra.
Under the influence of the cooling laser, the phonon occupation can be expressed as~\cite{aspelmeyer2014cavity} $\langle n(C) \rangle = n_{\mathrm{th}}[  \Gamma_0/\Gamma_+(C)] = n_{\mathrm{th}}/(1 + C)$, where $n_{\mathrm{th}}$ is the occupation of the phonon mode at thermal equilibrium. 
The thermal population, $n_{\mathrm{th}}$, is taken to be 22.4, consistent with an in-situ measurement of the \textmu HBAR temperature (13.6~K). Using $\Gamma_0$ and $\Gamma_+(C)$ derived from OMIT and spontaneous lineshapes, respectively, we evaluate this expression to find the phonon occupation versus cooling power, as shown in Fig.~\ref{fig:Cooling_exp}e.

Alternatively, the phonon population can be estimated based on the AS spectrum area (i.e. AS photon flux), using the expression $\langle n(P) \rangle = n_{\mathrm{th}} [ \tilde{\mathcal{V}}(P)\Gamma_+(P)/4\tilde C(P)\Gamma_0]$. 
Here, $\tilde{\mathcal{V}}(P)$ is the AS peak brightness normalized to 1 at $C=1$; $\Gamma_+(P)$ is the measured spontaneous linewidth;  $\tilde C(P)$ and $\Gamma_0$ are the cooperativity as a function of transmitted power and the fundamental linewidth, respectively, both extracted from the OMIT measurements.
As seen in Fig.~\ref{fig:Cooling_exp}e, the estimated phonon population based on the spectral area (diamonds) shows good agreement with that from the measured linewidth (circles); these data are also consistent with the trend line (dashed) obtained by evaluating $\langle n(P) \rangle = n_{\mathrm{th}} /(1 + \tilde C(P))$ using the power dependence of the cooperativity estimated from OMIT measurements. See Supplement Sec. IV for derivations of the above quantities and details of data analysis. 

These data demonstrate laser cooling of the \textmu HBAR to the ground state.
At the highest cooling power (1.3~mW), a phonon occupation of $< 0.4$ is reached; the phonon occupation is estimated from the AS spectra using the spectral linewidth (circles) and area (diamonds), yielding an occupation of $0.35\pm0.034$ and $0.375\pm0.038$, respectively.
Note that agreement between the theoretical and experimental phonon occupation seen in Fig.~\ref{fig:Cooling_exp}e is achieved at all control powers without a correction to the bath temperature, indicating an insignificant degree of absorption induced heating through these experiments.

To prevent unintentional heating of the phonon mode by laser noise during these measurements, a high-rejection (60~dB) filter was used before the cavity, to suppress any spontaneous photons and reduce phase noise at 12.6~GHz offset frequencies. 
Estimates based on the measured laser noise levels indicate that any heating caused by residual laser noise has a negligible effect, adding fewer than $2\times10^{-4}$ phonons to the phonon population at the maximum power levels used
(See Supplement Sec. VI-VII for details).
Hence, we have demonstrated laser cooling of such long-lived \textmu HBAR modes to their ground state, a critical milestone opening the door to the use of such systems as a quantum resource. 

\section*{Discussion}
We have presented a novel cavity optomechanical system that permits quantum optomechanical control of individual high coherence phonon modes supported by the \textmu HBARs for the first time.
It is important to note that no appreciable parasitic heating was observed through laser cooling of this system to the ground state, demonstrating improved robustness to problematic sources of decoherence that otherwise hinder efficient, low noise quantum operations.
Hence, such \textmu HBARs hold promise as the basis for a new class of quantum optomechanical systems with long coherence times and enhanced robustness to optical heating, addressing a critical challenge facing the field of quantum optomechanics~\cite{barzanjeh2022optomechanics,maccabe_nano-acoustic_2020}.

Looking ahead, the coherence times (Q-factors) of such \textmu HBARs are readily extended to 1.8ms ($140 \times 10^6$), using resonator geometries that produce lower surface participation~\cite{royce2024}. These enhanced coherence times and the improved robustness of such systems to heating could enable the realization of heralded single-photon sources~\cite{maggie2024} as well as high-fidelity quantum repeaters~\cite{fiaschi2021optomechanical}. Additionally, quantum optomechanical control of such \textmu HBARs having motional masses at or near the Planck mass (21.8 \textmu g), could also enable macroscopic tests of quantum coherence and backaction in regimes where both gravitational and quantum effects are important~\cite{galliou2013extremely,goryachev2014gravitational,lo2016acoustic,neuhaus2021laser,schrinski2023macroscopic,agafonova2024lasercooling1milligramtorsional}, providing new possibilities for innovative quantum sensors and fundamental tests of quantum mechanics in the macroscopic regime.

\vspace{10pt}

\section*{Acknowledgments}
We thank Dr. Yogesh Patil for helpful discussions of cryogenic system design.
Primary support for this research was provided by the U.S. Department of Energy (DoE), Office of Science, National Quantum Information Science Research Centers, Co-design Center for Quantum Advantage (C2QA) under contract No. DE-SC0012704. We also acknowledge supported by the Air Force Office of Scientific Research (AFOSR) and the Office of Naval Research under award No. FA9550-23-1-0338 and the National Science Foundation (NSF) under award No. 2137740. HHD acknowledges support from the Fulbright Israel program. Any opinions, findings, and conclusions or recommendations expressed in this material are those of the author(s) and do not necessarily reflect the views of the DoE, AFOSR, or NSF.

\section*{Data availability}
{The data supporting the findings of this study are available from the authors upon reasonable request.}

\section*{Author contribution}
{PTR, HHD and DM conceived and planned the experiments. HHD carried out the cavity optomechanical experiments, analytical derivations, and data analysis with assistance from YL, DM, TY, TBK and SG. YL fabricated the \textmu HBAR and performed spectroscopic measurements. DM, MP, RB, SP and JGEH contributed to the interpretation of the results. HHD and PTR lead the manuscript  writing. All authors provided critical feedback and helped shape the research, analysis and manuscript.}

\section*{Competing interests}
The authors declare no competing interests.

\bibliographystyle{naturemag}
\bibliography{main.bib}

\end{document}